# A Reliable And An Efficient Web Testing System

Kamran Ali and Xia Xiaoling

School of Computer Science and Technology, Donghua University,
Songjiang District, Shanghai 201620 – China

*ABSTRACT*

*To improve the reliability and efficiency of Web Software, the Testing Team should be creative and innovative, the experience and intuition of Tester also matters a lot. And most often the destructive nature of Tester brings reliable software to the user. Actually, Testing is the responsibility of everybody who is involved in the Project. But, one's personal curiosity and attention is more important than the various techniques and tools available in the market for Web Testing due to the phenomena that Software Testing is an art. In this study, we are actually discussing certain techniques and tools which can be helpful to minimize bugs in Web Application and achieve reliability and efficiency to a certain level. Indeed, for bettering the quality of Web Application, Testing may not be considered as the only effective method because no one can certify that a system is bug-free. This paper presents some essential web testing techniques, strategies, methods and tools which need to be focused on when performing Web Testing for several web applications in order to achieve better results.*

## 1. INTRODUCTION

In Software Testing, Check-out is the jargon that works everywhere in all activities of Software Life cycle. Though we are familiar that Testing is the fourth phase of Software Development Life Cycle but approximately 70% of development time is consumed on testing. As it is evident that Testing starts from requirement analysis phase and moves on till the last maintenance phase. Here as the requirement analysis and design phase starts, we also start applying techniques. The first technique that works here is to perform Static Testing. In Static Testing, we first of all evaluate Software Requirement Specification document whether it is according to the user requirements or not. Then, we perform deeper Static testing by looking at code reviews, code inspections, walkthroughs, design document and software technical reviews. As we move through Static Testing, the second technique, we use at this moment is Dynamic Testing. Dynamic Testing starts as the code is ready or just even a unit or module is ready. Dynamic testing utilizes three major testing techniques like Black box, white box and grey box testing for the assessment of code. One fact here we will discuss that though the developers build the code and indicate the absence of errors in code but despite this fact, errors come out. This is the fallible nature of human beings.

For the effective results, it is required for the Tester that he should have competency in testing techniques and undivided knowledge about the Application.that is under test. unit or module is ready. Dynamic testing utilizes three major testing techniques like Black box, white box and grey box testing for the assessment of code. One fact here we will discuss that though the developers build the code and indicate the absence of errors in code but despite this fact, errors come out. This is the fallible nature of human beings. For the effective results, it is required for the Tester that he should have competency in testing techniques and undivided knowledge about the Application that is under test.

This study discusses certain parameters that will help us in learning that why do we perform testing and what is its purpose and how it affects the reliability, quality and efficiency of web software [1].

- In the world of inter-connected computerized devices, the reliability and quality of Software is a matter of life and death and it can only be accomplished if we conduct extensive and rigorous testing.
- In the technical perspective, we can say that even the competent programmers are not infallible, most of the times; the implications of requirements are not foreseeable.
- Sometimes the behavior of the system cannot be forecasted from its components. Sometimes the languages, databases, operating systems and user interfaces have issues that cause trouble in the work of application.
- In the business case, if you do not find the errors, customers or users will.
- In the economic terminology, post-release debugging is considered as the most expensive form of development.
- Erroneous software also affects sale and reputation of company.

As it is evident that bugs are introduced at every stage of software development life cycle. Therefore, it is crucial to remove them as early as possible; so huge economic losses can be avoided at later stages and efficiency can be achieved owing to systematic testing. Thorough and rigorous testing is key in achieving reliability.

## 2. BACKGROUND

In the real testing scenario, when the coding is completed, the tester checks out the software for various errors. He reports the errors to Project manager as well as to Developers in the form of bug report. So, they can make necessary corrections and assess the proper working of Web Application. Sometimes even the non-serious errors are orally communicated by Tester to Developers and Project manager [2] so they can make required changes easily. Software testing is very hard; and particularly testing web applications is very onerous job.

Though Desktop systems resemble with Web Systems but the main characteristic of web application is its heterogeneous nature. Various operating systems, different hardware and network connections, multiple browsers and web servers, and the diverse use of technologies and programming languages make it more complex and testing such applications is a tedious task.

The wide diffusion of internet has also enhanced the importance of web application. But owing to market pressure, poorly documented code, time pressure, changing requirements, programming errors and miscommunication between client and project team, the sufficient time to the testing of web application is not provided that lowers the efficiency and reliability of web software. In this paper, we are presenting the problems and possible solutions which have been discussed by Researchers in the past two decades regarding web testing. Loopholes actually existed in all parts of software development life cycle. As we know that the basic techniques for web testing are almost same from the past two decades but they have now reached almost to its maturity. Web community is trying to offer possible solutions for achieving reliability and efficiency in web application. One fact regarding the web application is its distributive nature and access by multiple users under different platforms concurrently and it multi-tier architecture. Moreover, the interdependency in web system is also the major problem that affects the quality of web software.

Actually, there is certain essential criteria for measuring the quality of Web systems like functionality, reliability, usability, efficiency, maintainability, portability and productivity [3]. In our work, we are focusing on reliability and efficiency of web software by making deeper analysis of various web testing techniques, methods and tools in order to obtain satisfying results.

## 3. ESSENTIAL WEB TESTING STRATEGIES

It is observed that Systems are often deployed without being completely tested. But other fact is also evident that it is impossible to completely test the entire system. In that situation, we have to plan proper testing strategies so minimal amount of effort and time should be spent. Keeping all these facts in mind, we have to follow the testing strategies in this manner.

- First of all, we have to plan our test carefully. We must be well aware that what we are trying to test for.
- We must use the right test in the right situation like black box, white box or regression.
- We have to be systematic in debugging.
- The tester has also to ensure that what type of tests, he has to perform and how to perform them.
- Schedule of test also matters that when to run which test.
- For gaining efficiency in web testing, we have to evaluate our tests.
- We have to compare the results with the goals that we have set in Test Plan.
- We have to also prepare the metrics that let the Test Engineer determine the current quality of Web Software.
- For measuring the reliability of Web Software, the Tester has to assess the testing error rate of previous projects[4,25,26,27]

## 4. IMPORTANCE OF DOCUMENTATION IN WEB TESTING

Documentation plays a crucially important role in Web Testing. By documentation, we can properly estimate the test coverage, testing effort required and requirement tracking, etc. In this study, we are presenting certain documentation that is crucially important for most of the web applications to ensure quality to a certain level.

Table 1 :Documentation Artifacts

| | |
|---|---|
| Test Plan | It describes the testing strategies, resources and schedule |
| Test Case | It specifies one way of testing the system what to test with which input and result and under what conditions |
| Bug Report | It consists of list of bugs found by the Testing team for the software product that is under test. |

Documentation is a source to remove bugs because from all documentation, errors come out like Software requirement specification, design document, source code and test case so it is significant to first of all look for errors in them and fix them as early as possible [5]

**4.1 ESSENTIALS AND TECHNICAL ASPECTS OF WRITING SOFTWARE REQUIREMENTS SPECIFICATION**

Software Requirement specification document is the building block in software development. SRS plays a significantly important role for developing any type of Software. This is the first deliverable to the team for programming any sort of application like web application, mobile application and desktop application. After collecting all requirements from various stakeholders directly or indirectly involved in the project, the document that gets ready is Software Requirement Specification. This document contains overall description of a system that is to be developed. In software development, this document is the key before the start of coding and implementation of system. Everything regarding the system is defined at length in this document. SRS contains both functional and nonfunctional requirements of the system. We use activity diagrams, use case diagrams and data flow diagrams for providing ease to developers and sometimes we also insert snapshots and GUI screens, so, in future, possible changes can be made easily. Purpose and scope of Software is also discussed and possible constraints are also communicated to Client. This paper presents the key features and technical aspects of SRS document which help in resolving myriad problems which occur during designing and developing of Software Application. In this study, we will cover major areas of SRS document which will be of crucial help to whole project. Software Requirement Specification Document should be properly organized for the future project that is about to be designed so as the whole project team can extract all required information easily [5, 16].

First of all, feasibility study for that future project must be carried out in order to be aware of the fact that Software Project can transpire and be practically materialized.

After doing so, all technical areas of SRS should be properly worked upon. Early error localization and detection should be given undivided attention over other things. SRS should also be passed 0n from various testing phases so quality, efficiency and reliability for future Software System can be ensured.

Table 2 : Requirement Gathering Techniques

| Techniques | Description |
|---|---|
| Interviews | Client is usually a naïve user. Being a technical person, we conduct interviews in order to obtain technical information from all stakeholders. |
| Questionnaires | We sometimes use questionnaires to acquire data from clients. |
| Workshops | We also carry out workshops to take out hidden information from innocent clients. |
| Observations | Sometimes we get idea from previously designed projects that resemble with this expected project. |

### 4.2 DISCUSSION

SRS must be written in clear, precise and natural language. This is the guide to developers and whole project team. SRS is actually an agreement between customer and Software vendor. Cost estimation, budgeting, pricing and total effort of the project is estimated in SRS document. Deadlines for the project are also discussed into detail. In nutshell, we can say that everything regarding the project is jotted down in SRS document.

The most important thing in SRS is Project Title and then category of Software should be written whether it is Web, Mobile or Desktop Application. Existing and proposed system should be elaborated in detail. What will be the major benefits of proposed system which are missing in existing system? Software tools, deployment and hard specifications should be given for the proposed system.

Not understanding the clients requirements exactly and not documenting everything in advance will eventually give rise to overall cost and affect software life cycle. SRS is the parent document so every possible care must be must be made in order to avoid future anomalies. SRS document also reduces rework on project under development.

Omissions, misunderstandings, vagueness, incompleteness, human error as psychological or linguistic, and other technical errors should be avoided early in development life cycle by reviewing and assessing SRS meticulously. In recent years, the major work of Requirement Engineers is on fault localization and detection as early as possible in SRS, so, possible future effort and losses can be avoided. Proper requirement elicitation from various stake holders according to their wishes, needs and aspirations plays monumental role and converting these requirements technically by requirement analysts enhances the chance of success for the Project. There is also growing trend towards natural language processing to obtain requirements smoothly from naïve users [5, 17, 27, 26].

### 4.3 CLASSIFYING SOFTWARE REQUIREMENT ERRORS

Software Requirements early inspection and review by Project Managers is the important aspect that reduces overall effort at faster rate and thus developers can focus on other type of faults like missing or incorrect functionality which generate abruptly and human error that comes into the shape of requirements from client can be minimized[6,19].

### 4.4 FEASIBILITY STUDY

This study focuses on whether the Software Application can be practically materialized in terms of implementation and what can be possible constraints while developing this software. Feasibility report is a short document that discusses everything regarding the practicality of proposed Software.

### 4.5 A DEEP DIG INTO NON-FUNCTIONAL REQUIREMENTS

So far there in no unanimous consensus over non-functional requirements in Software Engineering community that how to gather, analyze and jot them down in organized manner whereas the significance of these requirements cannot be denied for the complete success of the Project. Indeed, these requirements are the body and soul of the system. For achieving quality attributes both functional and non-functional requirements must be taken into account. In the past, requirement engineers majorly focused on functional requirements whereas they undermined the importance of non-functional requirements which are the quality attributes of the System. In recent days, a great deal of work is being carried on which are the paramount quality attributes which must be dealt with properly for the minimal success of the Software System.

Table 3 : Technical Aspects of SRS

| Features | Technical Aspects |
|---|---|
| Consistent | There should be no clash or conflict in requirements at stage in whole SRS |
| Correct | What is stated is exactly the same that is desired by customer? |
| Unambiguous | There should be clarity in language of SRS |
| Complete | All functional requirements, non-functional requirements and constraints should be properly described. |
| Traceable | Source and origin of each requirement should be vivid and succinct. |
| Verifiable | Expected Software should be verifiable with SRS. |
| Modifiable | Changes could be made at any time in SRS while retaining its structure in original shape |

### 4.6 USAGE OF SRS

SRS is used by multiple people involved into the project.

**Customer**

When we handover the Software to client, he confirms the functionality of Software whether it is according to specified requirements in SRS or not.

**Software Developer**

Developer always writes code by looking at the requirements in SRS.

**Test Engineer**

Testing team also carries out testing of Software by following the requirements in SRS. So, helps greatly in removing bugs early in the project.

**Project Manager**

Project manager is the key person who while going through the SRS, divides the work among the project team for the expected software that is going to be built.

**Maintenance Engineer**

These people are not aware of the technical development of the software. They usually follow the SRS and provide the maintenance work because everything regarding the project is written in detail in SRS [6, 19, 20, 21].

### 4.7 EARLY ERROR DETECTION AND LOCALIZATION TECHNIQUES IN SOFTWARE REQUIREMENTS

Bug is the foundation stone of every problem that affects the quality and efficiency of Software. Most often majority of the errors are being generated in early phase of requirement elicitation from multiple stakeholders concerned with product in any way. Understanding the psyche and hidden mentality of client helps out significantly in minimizing mistakes in requirement gathering step. And human liability to error is also the key aspect which the developers must comprehend and take into account. In our paper, we are presenting certain error removal techniques which work out greatly to localize errors as early as possible in order to streamline the functionality and smoothness of Software and achieve reliability to a certain level. For instance the techniques of Information extraction, text mining, clustering techniques and analyzing the human textual knowledge by Requirement experts, bugs can be identified and removed early and thus quality can be ensured to some extent.

As it I evident that Developers besides understanding the technical aspects of Software, they must decipher the hidden meaning of requirements from clients. In simpler words, developers must improve their understanding of requirements as elicited by customers.

International Journal of Software Engineering & Applications (IJSEA), Vol.10, No.1, January 2019Research shows that understanding human cognition by developers can identify errors. Developers and whole project team must design certain methods which after incorporation can detect and localize errors and save crucial time which in response can be utilized on other development activities. Another technique for localization of errors is that certain models and datasets be designed which can translate and process the requirements as they are written in natural language. Moreover, programmers and clients must understand the relationship with each other about the product. Capability of Writer to formulate requirements and specify them properly ensures the quality in requirements and avoids lapses, mistakes and serious blunders in the long run in Software Development. Interviews, observations, questionnaires and other data gathering methods can be useful to remove bugs early in software requirements. If cognitive ability of customer is identified appropriately by Requirement Analysis's, it will strike enormously on the overall quality of Software System [6, 7, 20].

Table 4 : Error Detection Techniques in Software Requirements

| Techniques | Description |
| --- | --- |
| Aim | The objective of Software Application must be properly described. |
| Data collection | The methods for data collection should be stated by Requirement Analysts |
| Language Mastery | The writer for requirements must have good command over language in which he is jotting down requirements. |
| Previous Projects | the writer must look for similar and resembling projects previously written |

### 4.8 CASE STUDY

From the previous studies regarding the web systems, which we have discussed above, we come to know that majority of the problems occurred in all areas of software development life cycle as no required techniques, methods and strategies were used according to the needs of web software. Thus it affected the efficiency and reliability of web software. Therefore, in our study, we have endeavored to incorporate all the essential and crucial techniques in order to mitigate further anomalies.

The latest development in artificial intelligence, particularly in speech recognition system can mitigate the anomalies occurring in nonfunctional attributes and quality can be enhanced. Architectural erosion and poor traceability can only be avoided if we decipher and encode these requirements, specifically when we extract them from all stakeholders of the targeted system. Furthermore, for the smoother design of the software system, the identification and classification of these requirements is supremely important. In requirements prioritization major concern is to analyze the largest part of crucial requirements and their importance for a system. There are many methodologies, techniques that are used in requirements prioritization. All requirements cannot be enforced at same time, requirements are and have to be prioritized to be implemented to give

International Journal of Software Engineering & Applications (IJSEA), Vol.10, No.1, January 2019solution as early as possible in phases as scheduled in incremental design and evolution. These prioritization techniques are good for high quality software. In requirements prioritization, decisions are made by the stakeholders with respect to the product (attributes) and in contingency to external attributes such as time to market, deadlines, and usability thus it is important to prioritize requirements. The alignment of different software engineering activities for coordinated functioning and optimized product development is of great importance, particularly in industrial-scale development. The link between intermediate activities has been researched extensively, but the link between requirements engineering (RE) and software testing (ST) is a relatively less explored area. Several different techniques have been identified to improve RE and ST alignment. Test generation from requirements specification has received most attention. Alignment of RE and ST is particularly important for large safety-critical domains.

Table 5 : Error Localization Techniques

| Inspections | Requirement Engineers should look for errors early in SRS document |
|---|---|
| Reviews | There should be scheduled reviews by Audit team to avoid future anomalies |
| Overhauling | Project manager should personally investigate for any inconsistency, lapse or ambiguity in SRS and design document prior to code generation |

Customer involvement is the most important factor for the success of project and it is also obvious that requirement development is an inventive process, not simply a collection. Therefore, not only elicitation but also validation and management of requirements is pivotally important. Natural language limitations also cause problems in software requirements. Consequently, reading the mind of all stakeholders and writing expertise of Requirement Engineers helps out in locating and removing bugs early and are key to success in Software Projects [7, 21].

## 5. ESSENTIAL WEB TESTING METHODS: A DEEPER ANALYSIS

In this study, we are making a deeper comparison of three essential web testing methods like Black box, Grey box and White box testing methods [7, 8] so as the testers and developers can work out on the required methods according to the needs of the system.

International Journal of Software Engineering & Applications (IJSEA), Vol.10, No.1, January 2019Table 6 : Comparison of Testing Methods

| Black Box Testing | Grey Box Testing | White Box Testing |
|---|---|---|
| In this method, it is not required that the tester should be aware of the internal working of web software. | In this form of testing, the tester has limited knowledge of the internal workings of the web Software | In this method, the tester has the undivided knowledge about the internal working of the Application |
| This testing is also known as data-driven, closed box or functional testing | Also known as translucent testing, as the tester has limited knowledge of the insides of the application. | It is also known as code-based, structural or clear-box testing. |
| This testing is performed by developers, testers and also by end-users. | Performed by end-users and also by testers and developers. | It is most often done by developers and testers |
| This testing is based on the external behavior of the Web Application. | Testing is done on the basis of high-level database diagrams and data flow diagrams. | The tester can design and test the data accordingly because internal mechanism is known to him. |
| It is thorough and it takes less time | Partly time-consuming and exhaustive. | The most exhaustive and time-consuming type of testing |
| It is not suited for Algorithm testing. | It is not suited for Algorithm testing | It is suited for Algorithm testing |

## 6. IMPORTANT WEB TESTING TOOLS

As it is evident that a lot of automated tools are available in the market for Web Testing. But it is quite essential for the Tester that he should have mastery over various tools that are being used for web testing and then according to the needs of that Web Application, he should employ them and acquire the desired results within the minimal time frame and effort consumed. So, the proper evaluation and selection of every automated tool by the testing team for web testing also plays a significant role. Hundreds of the tools are available in the market [9, 12, 22] and released every year and each has its own limitations and benefits. There is one famous maxim that the more you understand the technology of your application, the better you will be able to test it. In this study, our focus is just to present a critical analysis of some well-known and popular web testing automated tools which can be considered helpful for several Web Testing Systems.

Table 7: Comparison of Software Tools

| Our Testing Criteria | Selenium | Ranorex | Test Complete |
|---|---|---|---|
| Execution speed | High | High | Very high |
| Cross platform | Windows, linux, unix, mac | Windows | Windows |
| Scripting language | Java, PHP, Python, Ruby, JavaScript, Perl, C# | No specific scripting lang. is used as it is written in .NET language using C#, VB.Net, and Iron Python | VBScript, Delphi, C++, C#, JavaScript |
| Data driven testing (Tools can accept and make changes in these without affecting test scripts) | Excel, Csv. | CSV, Excel, SQL | Excel, Csv, Sql |
| Playback capability | Supports record and playback option | Supports record and playback option | Supports record and playback option |
| Application support | Web Applications only | Web applications, mobile applications and desktop applications | Web app, mobile applications |

After looking into the technical aspects of these tools, we come to know that Ranorex is better one for Web-based Systems. Though each tool has its own merits and demerits but it solely depends upon the nature and features of application that the testing team has to decide to opt for which tool and how to obtain required results [9, 10, 11, and 22].

## 7. ESSENTIAL WEB TESTING TECHNIQUES: A DEEP LOOK INSIDE

As it is obvious that the requirements acquired from the client are either functional or non-functional in its nature. Similarly for assessing the quality of Web Software, we have to ensure the reliability and efficiency of Software in terms of both functional [11, 7, 9, 23] and non-functional testing in order to obtain better results.

Table 8 : Web Application Non-Functional Testing

| Testing Activity | Description |
|---|---|
| Reliability Testing | A test is considered as reliable if same result is obtained repeatedly. Reliability indicates the probability of failure free operation of Web Software for a specified period of time in a particular environment. |
| Compliance Testing | In this test, we verify that the system meets the timing constraints, usually for real-time and embedded systems. |
| Recovery Tests | In this test, we verify that the system can easily recover when it is forced to fail in multiple ways. |
| Volume Tests | We have to confirm that System can handle large amounts of data, high complexity algorithms or high disk fragmentation. |
| Compatibility Tests | By compatibility testing we can detect errors which occur owing to the usage of various client browsers and different Web Server Platforms. |
| Stress Tests | In Stress test, we execute the system beyond its normal conditions in order to verify whether the system crashes or it is able to recover from such conditions. We most often insert garbage data to assess such testing. |
| Accessibility Tests | In this testing our aim is just to access the web application with minimal hardware and software settings from the client side (such as browser settings disabling graphical visualization or scripting execution) or of users having physical disabilities (such as blind or deaf people). |
| Security Testing | Security testing is the major concern in Web Systems. The reliability and efficiency of web systems majorly depends on its security. In this test, unauthorized access by unauthorized users is checked out and monitored. And proper resources and services are delivered to authorized users. |

Table 9 : Web Application Functional Testing

| Unit Tests | Developers test the individual units of source code to determine whether they work properly or not. |
|---|---|
| Smoke Testing | This is the non-rigorous testing. We perform this testing to ascertain that important functions of the program work correctly. We don't go into finer details. |
| Sanity Testing | This testing offers shallow, broad and quick testing to determine whether it is reasonable and possible to proceed for further testing. |
| Integration Testing | In this test, we merge different units together and test collectively as a group to assess whether merged units or modules work together properly. |
| Regression Testing | After making new changes, we want to know whether older program still functions properly. |
| System Testing | Testing of the whole system to evaluate its compliance with the specified requirements. |

## 8. CASE STUDY

The history of past two decades tells us that a great of work has been done in Web Testing. According to bigger databases of Computer field such as ACM, IEEE, Springer and Science Direct, primarily 270 empirical studies regarding Requirement Engineering are identified and among them non function requirements is the least explored area. Today, the researchers are endeavoring to extract the true meaning and functionality of nonfunctional requirements and what actually is the definition of nonfunctional requirements will soon become evident before if we could analyze these quality attributes in true sense.

All the methods, techniques and strategies have almost remained same. These are now touching perfection to some extent. In Web Testing, nonfunctional testing has been the least explored area which is opening up now the new dimensions in the field of Requirement engineering. Some 360 quality attributes have been discussed by Researchers but our focus was to how achieve reliable and efficient web software by using minimal time and effort consumed. For doing this, we have imbibed all the essential techniques which can be of crucial help to achieve desired results. Understanding the real logic and phenomena of Software development life cycle is very important for testers and developers [24, 25, 26]. Tester should find the bugs and find them as early as possible and make sure that they are fixed.

## 9. DISCUSSION

Previously, in the past two decades a great deal of work has been done on web testing. But one thing that carries weight is that 'not selected and required techniques and tools' according to the needs of Web software are used. Thus, a lot of time and effort is consumed in wrong direction [13, 14, 15]. In our study, we have incorporated all the necessary techniques, tools and strategies for web soft wares, so, it would be easier for Testers to apply the required testing on web soft wares and avoid performing unnecessary operations and save the time and effort. In this way, reliability and efficiency can be obtained [27, 28]] in Web soft wares. Furthermore, in this way right test cases can be used, and implementation of proper automated tools will also be employed. Thus lofty losses can be avoided in the long run.

## 10. CONCLUSIONS

In this paper, our focus was on to achieve reliability and efficiency [14, 4] in Web Soft wares by using proper methods, techniques, strategies and tools of testing in order to obtain required results. And this can only be done if first of all we look deeply into the nature of Web Application then we apply and follow the essential testing rules mentioned above. In this study, the main conclusion is that SRS, Design document and Source Code are the key areas to remove bugs early and for achieving reliability and efficiency of Web software, proper and required strategies, techniques and tools should be adhered in order to get optimal results. Everybody who works in the development and testing of web software plays a key role in its reliability and efficiency [29]. Therefore, best results can only be acquired if everyone puts in their effort in the right direction. In my work, I have mentioned the merits and demerits of tools, thus, right choice of tool is vitally important as well. Therefore, huge budget losses, time and energy can be saved and effort should be put in the right place and in this manner overall reliably and efficiency can be bettered.

One thing that also carries weight in our research is that non-functional testing [30, 31] has not been given full attention, though the dimensions of efficiency and reliability are incomplete without proper nonfunctional testing. I have discussed above the major nonfunctional testing areas which need to be covered in order to obtain better results for web software that is the body and soul of web software.

International Journal of Software Engineering & Applications (IJSEA), Vol.10, No.1, January 2019International Journal of Software Engineering & Applications (IJSEA), Vol.10, No.1, January 2019

**AUTHORS**

**Kamran Ali Memon**, He received his Bachelors in Software Engineering from Mehran University of Engineering and Technology, Pakistan in 2009. He has been working as Senior Lecturer in Department of Software Engineering in Baluchistan University of IT, Engineering and Management Sciences, Pakistan. Currently, He is pursuing Masters of Engineering in Computer Science and Technology at Donghua University, Shanghai, China. His area of interest is Software Engineering, Software Testing and Software fault.

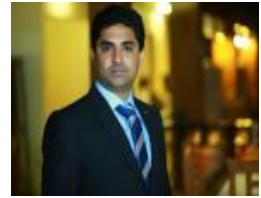

**Dr. Xia Xiaoling,** She received her Ph.D. in Computer Science from Shanghai Jiaotong University, Shanghai, China in 1994. Since then she has been working at Donghua University. At present she is Associate Professor in the School of Computer Science and Technology, Donghua University, Shanghai, China. Her area of interest is Artificial Intelligence, Big Data and Software Engineering.

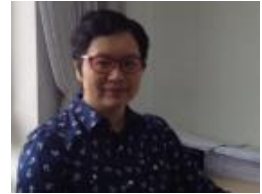
16